\documentstyle{l-aa}
\begin{document}

\def\ltsm{\mathrel{\raise.3ex\hbox{$<$}\mkern-14mu
                \lower0.6ex\hbox{$\sim$}}}
\def\gtsm{\mathrel{\raise.3ex\hbox{$>$}\mkern-14mu
                \lower0.6ex\hbox{$\sim$}}}

   \thesaurus{03
              (11.03.4 Cl~1613+3104;
               11.03.4 Cl~1600+4109; 
               11.16.1; 
               11.19.5)}

   \title{The population of galaxies in the distant clusters
Cl~1613+3104 and Cl~1600+4109\thanks{Based on observations made with 
the 3.5~m telescope of the Cen\-tro As\-tro\-n\'o\-mi\-co 
His\-pa\-no--Ale\-m\'an
de Ca\-lar Al\-to in Al\-me\-r\'\i a (Spain), 
operated by the Max-Planck-Institut 
f\"ur Astronomie, Heidelberg (Germany), jointly with the
Spanish Co\-mi\-si\'on Na\-cio\-nal de As\-tro\-no\-m\'\i a}}

   \author{R. V\'\i lchez--G\'omez\inst{1}
\and R. Pell\'o\inst{2}
\and B. Sanahuja\inst{1}
        }

   \offprints{R. V\'\i lchez--G\'omez}

   \institute{Departament d'Astronomia i Meteorologia, Universitat de
Barcelona, Av.\ Diagonal 647, E--08028 Barcelona, Spain
\and
Laboratoire d'Astrophysique de Toulouse, URA 285, Observatoire
Midi--Pyr\'en\'ees, 14 av.\ \'Edouard Belin, F--31400 Toulouse, France
             }

\date{Received 1995; accepted  1996}
\maketitle

\begin{abstract}
We study the photometric properties and content of the
distant clusters of galaxies Cl~1613+3104 ($z = 0.415$) and Cl~1600+4109
($z = 0.540$). The former is a rich, concentrated cluster which shows a
strong evidence for segregation in luminosity and color. Its population of
galaxies separates into two different kinds of objects: a red population
compatible with the colors expected for 
E/S0 galaxies located preferentially
in the central part of the cluster, and a sparse blue population 
explained in part
by the presence of normal S/Im galaxies, which is much less concentrated. 
In the case of Cl~1600+4109, there is a lack of E/S0 galaxies and the population
is mainly made up of S/Im galaxies, without any evidence of segregation in
magnitude or in color. Both clusters
exhibit an important fraction of blue objects, increasing with magnitude. 
This result is due in part to the presence of normal S/Im galaxies,
and also to extremely blue objects probably undergoing an episode of intense 
star-formation.

\end{abstract}

\section{Introduction}

   Photometry of rich clusters of galaxies at high redshift
remains one of the most important tools to understand the evolution of
galaxy populations in dense environments. The discovery of a large fraction of
blue galaxies in rich clusters has
been understood as evidence for a strong, recent evolution of galaxies
in clusters (Butcher \& Oemler 1984, hereafter BO; Sharples et al.\ 1985).
%% This is known as the ``Butcher--Oemler effect''. 
The interpretation
of this blue excess is a difficult challenge because the galaxy
content varies from cluster to cluster. 
Dressler et al.\ (1985) suggested that it 
can arise from several reasons: different initial
conditions, a difference in the time scales for cluster formation and/or
environmental influences. The presence of luminosity and color segregation
in nearby clusters (Capelato et al.\ 1980) as well as in
high redshift clusters (Mellier et al.\ 1988) is often interpreted in terms
of dynamic friction and environmental influence (galaxy-galaxy and
galaxy-ICM interactions). Nevertheless, it is not still clear
to what extent these properties are innate or the result of later
evolutionary processes. Dressler \& Gunn (1992) published
a paper on the photometry and 
spectroscopy of seven clusters in the redshift range $0.35 \leq z \leq 0.55$,
including a thorough discussion on the statistical properties of the sample. 
Such studies are crucial to an understanding of
the evolutionary processes that take place in clusters. 
Recently, Dressler et al.\ (1994a,b) and Coutch et al.\ (1994) studied
the morphology of the blue galaxies in clusters at $0.3\ltsm z \ltsm 0.4$, using
HST images. They found that the blue objects are predominantly disk-dominated
or irregular galaxies, involved in merger or interaction processes.
Thus, it is important to enlarge the sample of 
well known high and medium redshift
clusters, and photometry allows to study a larger sample of galaxies
than spectroscopy. 

Cl~1613+3104 ($\alpha_{1950}=16^{\rm h} 13^{\rm m} 49\fs2$,
$\delta_{1950}=31\degr 4\arcmin$ $57\farcs0$; $l=50\fdg58$, $b=+45\fdg56$) 
is a mid-redshift rich cluster, which exhibits X-ray (Henry et al.\ 1982) and
radio emission (Jaffe 1982). Its redshift ($z=0.415$) was obtained 
by Sandage et al.\ (1976) from the  spectrum of the first-ranked galaxy. 
V\'\i l\-chez--G\'o\-mez et al.\ (1994) reported the
existence of a diffuse light extending up to 265 h$^{-1}$ kpc
from its center, associated with a stellar component immersed in the
gravitational potential of the cluster.
Cl~1600+4109 ($\alpha_{1950}=16^{\rm h}~0^{\rm m}~23\fs0$,
$\delta_{1950}=41\degr~9\arcmin~39\farcs0$; $l=65\fdg21$, $b=+48\fdg72$) 
is another mid-redshift cluster ($z=0.540$; Henry et al.\ 1982) 
included in the sample of Gunn et al.\ (1986),
whose population of galaxies has not been studied to date. It is not
a strong X-ray emitter: its upper limit for the X-ray emission is
$1.6\times10^{44}$~erg~s$^{-1}$ in the 0.5--4.5~keV range (Henry et al.\ 1982).
The database used here comes from 
the photometric survey by Pe\-ll\'{o} \& V\'{\i}l\-chez--G\'{o}\-mez
(1996), hereafter referred as Paper I.

    The layout of this paper is the following. A summary of
the observational procedure is presented in Sect.\ 2. Section 3 reviews the
methods used in the analysis of the photometric catalogs and the results
obtained are given in Sects.\ 4 and 5.
In Sect.\ 6 we discuss these results and we present the  
conclusions of this work. We assume
$H_{0}=50~{\rm km~s}^{-1}~{\rm Mpc}^{-1}$ and $q_{0}=0.1$ in a standard
Friedmann cosmology.

\section{Summary of the photometric observations}

    The observations reported here were carried out in two different
runs (in May 1987 and July 1988) at the F/3.4 prime focus of the 3.5~m
telescope of the {\it C.A.H.A} (Ca\-lar Al\-to, Al\-me\-r\'\i a, Spain). 
All the details can be found in Paper I.
The detector used was an RCA CCD, with an equivalent field of
$4\farcm4\times2\farcm8$ and a pixel size of 0\farcs506. 
The filters used were Johnson B and Thuan--Gunn g and r 
(Thuan \& Gunn 1976), with effective wavelengths of 4416, 4892 and 6681~\AA, 
respectively. These filters 
avoid the most important emission lines from the sky, which produce
a strong fringing in RCA CCDs. As a consequence, the S/N of the images is
improved and the detection level is much better. A more important reason to 
select these filters refers 
to the redshifts of both clusters and the predictions obtained
by the synthetic models of spectro-photometric evolution
(Bru\-zual \&  Charlot 1993;  Rocca--Volmerange \&  Guiderdoni 1988).
The magnitudes derived from these filters refer to fluxes emitted at 3120, 
3457 and 4721~\AA\ in the rest frame of Cl~1613+3104, 
and at 2867, 3177 and 4338~\AA\ in the rest frame of Cl~1600+4109. 
The main feature in the spectrum of galaxies in the optical range is the
discontinuity at 4000~\AA, which is located between filters g and r for
galaxies at these redshifts ($z\sim0.4$--0.5). Thus, the ${\rm g}-{\rm r}$ color
gives a good measure of the relative spectral flux before and after the
discontinuity, and it is a useful tool to discriminate between cluster and 
field galaxies. Furthermore, for the same redshift range,
the B filter samples a region of the spectrum very 
sensitive to star formation activity. These filters allow
to obtain some reliable information about the population of
galaxies and the star formation history in these clusters.  

For  each  filter  and  cluster,  we  took equivalent exposures on an
empty field  near the cluster ($10\arcmin$ southward from 
the center of Cl~1613+3104, and  $7\arcmin$ southward and $7\arcmin$ 
westward from the center of Cl~1600+4109). 
These images were reduced in the same
way as the cluster ones. The comparison field allowed the statistical
correction of the cluster content for the contamination due to
foreground and background objects within the same line-of-sight. 
Table 1 of Paper I gives the main characteristics of the photometric CCD
images of the clusters and the respective comparison fields used here.
To be sure that the comparison fields are
not contaminated by the presence of other clusters of galaxies, we used
Zwicky et al.'s catalog (1961--1968) to 
determine  the  position  of  the  comparison  field with respect to the
cluster. An estimate of the foreground reddening was obtained using
the maps of Burstein \& Heiles (1982) and the Galactic extinction
law (Cardelli et al.\ 1989). The reddening correction to each color is
less than 0.01 mag in all the cases, which is consistent with 
the colors obtained for objects identified as stars (Paper I). Thus, 
we did not introduce the reddening correction in our analysis because
it is much lower than the absolute photometric 
accuracy for the brightest objects in the field.

\section{ Data analysis }

    The first goal is to estimate the distribution of galaxies in 
spectro-morphological types through the observed distribution in colors.
We used Bruzual's code (Bruzual \& Charlot 1993) for 
the spectro-photometric evolution of galaxies,
in order to obtain the SEDs corresponding to 6 spectro-morphological types:
E/S0, Sa, Sbc, Scd, Im and a single starburst. The IMF was taken from
Miller \& Scalo (1979), with a lower and upper cutoff masses 
of 0.1 and 125.0 $M_{\odot}$ respectively. For each morphological type,
we calculate the evolution of colors as a function of the redshift.
The present age of galaxies is assumed to be 15 Gyr, which corresponds 
in this cosmological model to a galaxy formation redshift of 5.3.
The real transmission functions of the photometric system constituted by the
filters plus the CCD response are taken into account
to derive the k and e corrections to magnitudes and 
the synthetic colors. In this
way, observations can be easily compared to color predictions and
undesirable color effects are avoided. 
The fraction of blue objects, with a definition which is similar to
that given by BO, is also determined.

%\begin{figure}
\begin{figure*}
%\picplace{9cm}
\picplace{1cm}
%\centerline{\hbox{ \psfig{figure=fig1a.ps,height=11.6cm,width=6.0cm}
%\psfig{figure=fig1b.ps,height=11.6cm,width=6.0cm}
%\psfig{figure=fig1c.ps,height=11.6cm,width=6.0cm}}}
\caption []{Distribution in absolute B magnitude and colors of objects in the
field of Cl~1613+3104, before (white histograms) and after (black)
correction for contamination using the comparison field:
{\bf a} Isophotal B magnitude,
{\bf b}  ${\rm B}-{\rm r}$, {\bf c}  ${\rm g}-{\rm r}$. In b and c,
only objects brighter than the completeness magnitude in g have been
considered. $\pm 1 \sigma$ error bars at the bottom of the histograms
correspond to the net corrected population
}
\end{figure*}
%\end{figure}

    The contamination due to fore and background galaxies has
been evaluated as a function of the different magnitudes and colors
by means of the comparison field. This is a key point because 
the population of cluster galaxies appears as an excess
with respect to the field population. In particular, 
the Schechter luminosity function (Schechter 1976) is used to fit
the  distribution of galaxies in B,
$$\vbox{\halign{# \cr $\displaystyle{
{\rm n} (L) {\rm d} L = {\rm n}^{\ast} 
\Biggl({L \over { L^{*}}} \Biggr)^{\alpha}
{\rm e}^{ - L / L^{*}} {\rm d} \Biggl({L \over { L^{*}}} \Biggr) {\rm ,}}$ \cr
}} \eqno (1)$$
through a Gauss--Newton $\chi^2$ minimization algorithm. 

    The 2D distribution of galaxies has also been studied.
To evaluate the degree of central concentration of the clusters,
we computed the concentration index (C), R$_{30}$ and N$_{30}$, as defined 
by Butcher \& Oemler (1978). Nevertheless, these parameters need an 
accurate determination of the extent of the cluster and a good
correction of the fore and background contamination. Therefore, we also 
used the average central density, N$_{0.5}$ (Bahcall 1981), which is 
very easily determined and almost insensitive to these errors.
%% on the position of the cluster center or to the background correction.

    The projected densities of these clusters (number of galaxies per surface 
unit or surface brightness) were fitted using a Gauss--Newton 
$\chi^2$ minimization algorithm to the analytic profile given by:
$$\vbox{\halign{#   \cr
$\displaystyle{   \sigma(r)=\sigma_0\Biggl(1+\biggl({ r\over{\rm
R}_{\rm c}}\biggr)^2\Biggr)^{-\alpha}{\rm .}}$ \cr
}}  \eqno (2) $$
This law corresponds to the Hubble profile when $\alpha = 1$.
Unless otherwise indicated, the sample used is
defined within the completeness limit in magnitude.
The center of the cluster is defined as the 
optical barycenter of the central brightest galaxy, when there is only one 
(Cl~1600+4109), or as the optical barycenter of the whole central
group of galaxies (Cl~1613+3104). 

    The existence of subclustering and
the presence of segregation in luminosity and color can be detected performing
the angular-separation test ($\lambda_{\rm n}$-method) 
proposed by Hickson (1977), as developed by Capelato et al.\ (1980). 
This method introduces a set of length scales
defined by:
$$\vbox{\halign{# \cr $\displaystyle{
\lambda_{\rm n}=\biggl({1\over{ N_{\rm p}}}\sum_{i>j}
r_{ij}^{\rm n}\biggr)^{1/{\rm n}}{\rm ,}}$ \cr
}} \eqno (3)$$
where $N_{\rm p}$ is the number of pairs of galaxies and  $r_{ij}$,
the angular separation between galaxies {\it i} and {\it j}. 
The $\lambda_{\rm n}$ values are calculated 
over successive magnitude or color
intervals, to check for the existence of segregations in luminosity  
and color. This test has the advantage of being independent of the
position of the center and the shape of the cluster.
%% (no circular symmetry is needed).
We adopted the definition of {\it segregation} given
by Capelato et al.\ (1980).
%% a group of galaxies, which represents an
%% interval in magnitude and/or color, is segregated when its
%% values of $\lambda_{\rm n}$ are smaller than those of any other group. 
We used $\lambda_{1}$ and $\lambda_{-1}$ and the
field has been limited within a circle smaller than the
image size to avoid the edge effects. The size of
the different magnitude or color intervals were imposed 
in such a way that each one contains approximately the same number
of galaxies.

\section{ Cl~1613+3104 }

\subsection{ The population of galaxies }

Figure 1 shows the histograms in absolute magnitude for the B filter
as well as the distribution in colors, before and after correction
for contamination using the comparison field. Only objects
unambiguously identified as stars have been excluded from the sample
(following the method explained in Paper I). 
The luminosity function in B  has been derived according to Eq.\ (1),
using the same k correction for all the objects 
(assuming that they are all E/S0 galaxies). 
The fit gives: ${\rm n}^{\ast}=40\pm20$, 
$\alpha=-0.8\pm0.6$ and 
${\rm M}^{\ast}_{\rm B}=-20.8\pm0.8$ (reduced $\chi^2$ is
3.75, with 3 degrees of freedom). These values are compatible
with those derived by Colles (1989) for a 
sample of 14 rich and nearest ($z<0.15$)
clusters, also using a $\chi^2$ minimizing method.
If we assume $\alpha\equiv-1.25$, then,
${\rm n}^{\ast}=19\pm6$ and ${\rm M}^{\ast}_{\rm B}=-21.4\pm0.5$
(the reduced $\chi^2$ is 1.78 with 4 degrees of freedom),
which are very close to the values obtained by Colles, fixing also $\alpha$.
The distribution of galaxies in colors clearly shows a double peak 
in ${\rm g}-{\rm r}$. This configuration divides the sample of
galaxies into blue ($0.0\ltsm{\rm g}-{\rm r}\ltsm0.7$) and red
($1.0\ltsm{\rm g}-{\rm r}\ltsm1.5$) populations. The separation into
populations is less evident in ${\rm B}-{\rm  g}$, which shows only one
broad peak centered around 0.4. In ${\rm B}-{\rm  r}$, 
the sample of galaxies shows an important fraction of blue
galaxies (${\rm B}-{\rm  r} \ltsm 1.8$) and the presence of a
red population  (${\rm B}-{\rm  r} \gtsm 2.0$).

Figure 2 shows the color--magnitude
diagrams for the cluster, where the separation between stars, galaxies
and unidentified objects has been introduced according to 
Table 2 in Paper I. The red sequence 
for the distribution of E/S0 galaxies is less relevant
than in other clusters (for example, compared with BO or Dressler
\& Gunn (1992), for the same type of clusters). An
important blue population appears in these diagrams, as preliminary evidence
in favor of an important fraction of star--forming galaxies. This population 
is examined below.

\begin{figure}
%\picplace{8cm}
\picplace{1cm}
%\centerline{\psfig{figure=fig2.ps,height=11.6cm}}
\caption []{Color--r magnitude diagrams in the field of Cl~1613 +3104:
{\bf a} $({\rm B}-{\rm r})$,  {\bf b} (${\rm g}-{\rm r}$).
The (B $-$ r)(r) locus of B$_{\rm com}$ and the (g $-$ r)(r) locus of 
g$_{\rm com}$
are given by solid lines, whereas r$_{\rm com}$ is shown by a dashed line.
Galaxies are plotted as black dots, whereas stars and
unidentified objects appear as stars and plus signs, respectively
}
\end{figure}

Table 1 gives the distribution in spectro-photometric types of objects
in excess in the cluster field with respect to the comparison field.
Spectrophotometric types are determined as boxes on the color-color plane.
Extremely blue objets (EB) are objects bluer than 
the values predicted for an Im type at the cluster redshift.
%% (they belong to the blue peak in the g $-$ r 
%% distribution shown in Fig.\ 1). 
Again, only objects identified unambiguously 
as stars have been excluded from the sample.
Figure 3 shows the color-color distribution of objects for three
intervals in r magnitude, where we superimposed the synthetic sequence
of morphological types derived using Bruzual's code. 
The population of blue galaxies in the color--color diagram increases 
with the magnitude,
as shown in Table 1. Nevertheless, to obtain the absolute distribution in 
spectro-morphological types 
from these values, the sample must be limited within the
magnitude of completeness for {\it all} 
the filters (see discussion in Sect.\ 6).

\begin{table}
\caption[]{Distribution of the excess of galaxies in the field of
Cl~1613+3104 with respect to the comparison field. }
\begin{flushleft}
\begin{tabular}{lrrrl}
\hline\noalign{\smallskip}
%\halign{#\hfil&&\quad#\hfil\cr
%\noalign{\hrule\medskip}
 & E/S0 & S/Im & EB \\
%\noalign{\smallskip}
%\hline\noalign{\smallskip}
\noalign{\medskip\hrule\medskip}
(B $-$ r) &$ [1.8, 2.9] $&$ [0.6, 1.8] $&$ < 0.6$\\
(g $-$ r) &$ [1.2, 2.0] $&$ [0.4, 1.2] $&$ < 0.4$\\
%\noalign{\smallskip}
%\hline\noalign{\smallskip}
\noalign{\medskip\hrule\medskip}
All & 27\% & 28\% & 45\% \\
${\rm r}\leq{\rm r}_{\rm com}$ & 44\% & 33\% & 23\% \\
${\rm r}\leq22$ & 75\% & 22\% & 3\% \\
%\noalign{\medskip\hrule\medskip}}
\noalign{\smallskip}
\hline
\end{tabular}
\end{flushleft}
\end{table}

All these data suggest the following
question: How do classify BO ``blue'' objects 
identified in this field according to this scenario? Following 
BO, a galaxy is ``blue'' when the 
${\rm B}-{\rm V}$ color in its rest frame is at 
least 0.2 mag bluer than the center of
the E/S0 peak, and the fraction of blue galaxies, ${\rm f}_{\rm B}$,
is obtained as the fraction of such galaxies with respect to the total
corrected population of galaxies, with M$_{\rm V}<-20$ 
within R$_{30}$. Then, the total number of objects with 
M$_{\rm V}<-20$ (equivalent to ${\rm r}<21.8$ for elliptical 
galaxies at $z = 0.415$) is only 60 within the cutoff radius of the cluster
(see Sect.\ 4.2), and 21 within R$_{30}$ (= 42$\arcsec$, Sect.\ 4.2). 
Eight of them have ${\rm B}-{\rm g}$ 
colors 0.2 mag bluer than a typical E galaxy. The equivalent values 
in the comparison field within R$_{30}$ are 8 and 3, respectively, so the
contamination level is about 40--50\%. The fraction of 
blue objects inferred in this way is ${\rm f}_{\rm B}=0.4\pm0.2$,
which is higher than the typical value and only marginally compatible with 
the {\it z}--f$_{\rm B}$ relation
obtained by BO for a rich cluster at $z=0.415$ (${\rm f}_{\rm B} \sim 0.2$).
This fraction increases if we redefine the 
fraction of blue objects by considering the
sample within the completeness limit in g (g$_{\rm com}$ = 24.25) and within
R$_{30}$. The total number of objects is 41, with a contamination level of
about 40 to $50\%$. Among them, 22 are ``blue'' objects and the
contamination level for this sample is also about $50\%$, so
${\rm f}_{\rm B}=0.47\pm0.19$. 14 of these objects are
compatible with stars according to their colors. The remaining objects
are classified as follows, assuming that they are cluster members: 4 as 
S galaxies, 3 as Im galaxies and 1 as an extremely blue object.

\begin{figure}
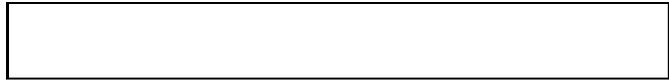
 
%\picplace{11.5cm}
\picplace{1cm}
%\centerline{\psfig{figure=fig3.ps,height=11.6cm}}
\caption [] { Color--color diagrams in the field of Cl~1613+3104 and
in its comparison field. Three magnitude 
intervals have been considered: {\bf a and b} ${\rm r} \leq 22$, {\bf c and d} 
$22 <{\rm  r} < 24$, {\bf e and f} ${\rm r} \geq 24$. Objects are identified
as in Fig.\ 2. The colors predicted for
the different types of galaxies at $z=0.415$, from E to Im, 
and the typical error boxes in each magnitude bin are also
shown
} 
\end{figure}

\subsection{ The 2D distribution of galaxies}
The shape of Cl~1613+3104 is basically circular with an ellipticity
$\epsilon\simeq0.15$ (defined as $\epsilon=1-\displaystyle{{\rm b}/{\rm a}}$,
where a and b are, respectively, the major and the minor axes
of the ellipse)
and a position angle $\theta\simeq100\degr$ with respect to the N--S 
direction (V\'\i l\-chez--G\'o\-mez et al.\ 1994).
This ellipticity is small enough to assume radial symmetry.
%% in the subsequent calculations.
The cluster core presents a compact group of 6 luminous galaxies   
(Fig.\ 1 in Paper I) within a radius of about $17\arcsec$ (121 h$^{-1}$ kpc)
around the center, with magnitudes between ${\rm r}= 19.13$ and
${\rm r}= 20.33$. 

\begin{figure}
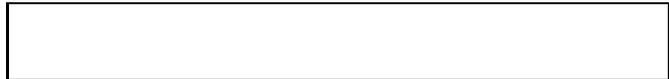
 
%\picplace{10.5cm}
\picplace{1cm}
%\centerline{\psfig{figure=fig4ab.ps,height=11.6cm}}
%\centerline{\psfig{figure=fig4cd.ps,height=11.6cm}}
\caption [] {{\bf a} $\lambda$ versus r magnitude diagram 
in the field of Cl~1613+3104. {\bf b}
$\lambda$ versus (${\rm g}-{\rm r}$) diagram. 
{\bf c} Magnitude segregation in two different color samples. 
{\bf d} Color segregation in three different magnitude samples.
All the bins contain, approximately, the
same number of objects}
\end{figure} 

Two partially overlapping images are available in r, one of them
shifted $71\arcsec$ with respect to the other. 
Therefore, as the spatial coverage is larger
in r, we have used these images to estimate the concentration of the cluster.
The equivalent image on the comparison field gives the
zero point for the projected density of galaxies. 
We have calculated the cluster profile and we have defined
the cutoff radius as
the distance at which the projected density coincides with the density in
the comparison field. The cutoff radius obtained is 
$101\arcsec$ (718 h$^{-1}$ kpc). The total number of objects inside
this radius is 456, and 207 of which have a magnitude lower than 
the completeness limit in r (r$_{\rm com}$). 
The number of objects found for the comparison field, under the 
same conditions, is 264 and 90, respectively. Assuming that all galaxies
are elliptical, 
the magnitude r$_{\rm com}$= 24.0 is equivalent to ${\rm M_{\rm V}}=-17.8$,
taking into account the distance modulus and the k+e correction.
With this limit of magnitude, the concentration parameter derived
is ${\rm  C}=0.36$, with ${\rm R}_{30}=41\arcsec$ (292 h$^{-1}$  kpc)
and N$_{30}=35$, indicating a highly concentrated cluster. If we
take the limit in magnitude used by BO (${\rm M_{\rm V}}= - 20.0$, 
${\rm  r}= 21.8$), we obtain a concentration parameter ${\rm  C}= 0.52$,
with ${\rm R}_{30}=42\arcsec$ (298 h$^{-1}$  kpc) and N$_{30}=13$. In 
this case, the total number of objects in the cluster field is much
lower: 60 compared to 16 in the comparison field. The central
galaxy density, ${\rm N}_{0.5}$, is 24 and the velocity dispersion, 
estimated from the empirical relation found by
Bahcall (1981), is 1000--1200~km~s$^{-1}$.
All these results confirm that we are dealing with a rich, compact cluster.

The projected galaxy density profile can be derived by fitting
the cluster profile according to Eq.\ (2):
$\sigma_0=0.015\pm0.004~{\rm gal~arcsec}^{-2}$, $\alpha=0.8\pm0.2$ 
and the core radius is R$_{\rm c}=21\arcsec\pm8\arcsec\, 
(150\pm60~{\rm h}^{-1}~{\rm kpc})$.
The radius at which the projected density reduces to $\sigma_0/2$
is  R$_{\rm h}=25\arcsec\pm10\arcsec\, (180\pm70~{\rm h}^{-1}~{\rm kpc})$.
These radii are similar to 
the typical radii found for other rich clusters with similar redshifts,
such as A~370 (Mellier et al.\ 1988). 

\begin{figure*}
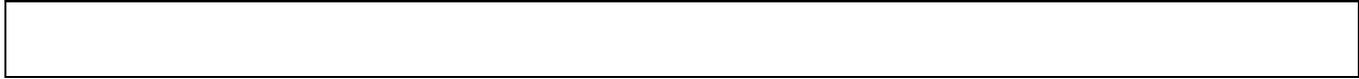
 
%\picplace{14.5cm}
\picplace{1cm}
%\centerline{ \hbox{\psfig{figure=fig5a.ps,height=12.6cm}
%\psfig{figure=fig5b.ps,height=12.6cm}}}
\caption [] { Isocontour maps of projected number density
of galaxies:
{\bf a} red galaxies ($1.0\ltsm{\rm g}-{\rm r}\ltsm2.0$); 12 contours 
levels are shown for 1 to 13 times the mean projected density over 
the cluster field, starting at 1.
{\bf b} blue galaxies ($0.0\ltsm{\rm g}-{\rm r}\ltsm1.0$); the 
difference between two successive contour levels is 0.5 times the mean
projected density over the cluster field, starting at 1. 
X and Y coordinates are in arcsecs with 
the cluster center at ($80\arcsec$, $130\arcsec$).
See comments in Sect.\ 4.2 }
\end{figure*}

   The luminosity and color segregations are important in Cl~1613+3104. 
Figure 4 displays the values of the characteristic lengths for
different magnitude and color bins. There is a segregation in
luminosity (manifest for $\lambda_{-1}$ in Fig.\ 4a) as well as a clear
segregation in color (Fig.\ 4b). To check on the independence of these 
segregation effects, we have applied this method to different
color and magnitude samples within the completeness limits in magnitude 
(Figs.\ 4c and 4d). The brightest and 
reddest objects are more concentrated because their characteristic
lengths are systematically shorter than those for the faint, blue objects.
In contrast, bright {\it and} blue objects appear anti-segregated,
as expect for stars and foreground galaxies in this small field.
If we select a sample of objects with the same $\lambda$, 
in the color range $0.75 \ge {\rm g}-{\rm r} \ge 1.8$, the 
luminosity segregation becomes evident (Fig.\ 4c). Conversely, a sample of
magnitude-selected objects with the same $\lambda$, 
shows clearly a color segregation (Fig.\ 4d). So, these segregation effects are 
independent, at least up to ${\rm r} \simeq 23$. Objects fainter than 
${\rm r} \simeq 23$ are quite uniformly distributed. Moreover,  
the red, luminous galaxies preferentially appear in the
central part of the cluster. If we consider the magnitude and color
distribution of objects brighter than r$_{\rm com}$
as a function of their radial distance to the central galaxy,
galaxies lying within a radius of $15\arcsec$ (107 h$^{-1}$ kpc) from the 
center are redder and brighter than the outside population
(the contamination expected in this region is lower than $20 \%$).
Figure 5 shows the contour plots of the projected number density of galaxies, 
for a red sample ($1.0\ltsm{\rm g}-{\rm r}\ltsm2.0$) and
a blue sample ($0.0\ltsm{\rm g}-{\rm r}\ltsm1.0$), 
according to the distribution of galaxies in ${\rm g}-{\rm r}$ (Sect.\ 4.1). 
The mean projected densities obtained over the cluster field for these 
two samples are $1.8\times10^{-3}$ and 
$5.0\times10^{-3}$ objects~arcsec$^{-2}$ respectively.
These values can be compared to the equivalent ones in the comparison
field: $0.6\times10^{-3}$ and  
$3.4\times10^{-3}$ objects~arcsec$^{-2}$ respectively. In conclusion, the 
red population is strongly concentrated around the 
center of the cluster (Fig.\ 5a)
whereas the blue is uniformly distributed (Fig.\ 5b). Very similar 
2D distributions are obtained by dividing 
the sample into bright (${\rm r} \le 23$)
and faint (${\rm r} \geq23$) objects, which are fully compatible with
the results shown in Figs.\ 5a and 5b for the red and the blue
populations, respectively.

\section{ Cl~1600+4109 }

\subsection{ The population of galaxies }
The number of objects detected in this field is lower than in Cl~1613+3104. 
334 objects were found in the cluster field compared to 
192 objects in the comparison field, so the excess of objects potentially 
belonging to the cluster is 142. For this reason, the analysis of
the population of galaxies in this case is more difficult than in
Cl~1613+3104. The distribution of these objects in absolute magnitude in B
(a k correction for E/S0 galaxies was assumed) and colors
${\rm B}-{\rm r}$ and ${\rm g}-{\rm r}$, before and after correction for
contamination, is shown in Fig.\ 6. Most of the objects belonging to the 
cluster are beyond the completeness limits in magnitude:
$24\ltsm{\rm B}\ltsm27$, $24\ltsm{\rm g}\ltsm27$ and $23\ltsm{\rm r}\ltsm27$.
A tentative fit of the luminosity function in B, according to Eq.\ (1),
gives ${\rm n}^{\ast}=12\pm2$ and ${\rm M}^{\ast}_{\rm B}=-22\pm4$,
with $\alpha\equiv-1.25$ (the reduced $\chi^2$ is 4.3, with 1 degree 
of freedom).

\begin{figure*}
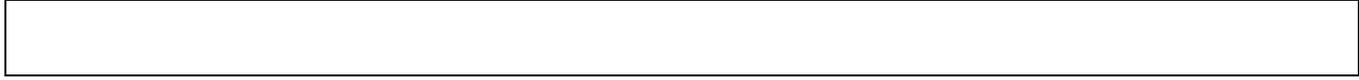
 
%\picplace{9cm}
\picplace{1cm}
%\centerline{\hbox{ \psfig{figure=fig6a.ps,height=11.6cm,width=6.0cm}
%\psfig{figure=fig6b.ps,height=11.6cm,width=6.0cm}
%\psfig{figure=fig6c.ps,height=11.6cm,width=6.0cm}}}
\caption []{ Distribution in absolute B magnitude and colors of objects 
in the field of Cl~1600+4109, before (white histograms) and after 
(black) correction for contamination: {\bf a} Isophotal B magnitude, 
{\bf b}  ${\rm B}-{\rm r}$, {\bf c}  ${\rm g}-{\rm r}$. $\pm 1 \sigma$ 
error bars at the bottom of the histograms correspond to the net corrected
population
}
\end{figure*}

The distribution of galaxies in colors takes into account all the objects
detected, without introducing any selection in magnitude. These
distributions, after correction using the 
comparison field, are rather irregular. 
Both the ${\rm B}-{\rm r}$ and the ${\rm g}-{\rm r}$ histograms show
a broad distribution of blue objects ($1.0\ltsm{\rm B}-{\rm r}\ltsm2.5$
and $0.2\ltsm{\rm g}-{\rm r}\ltsm1.6$), with very few red objects
(${\rm B}-{\rm r} \gtsm 2.5$ and ${\rm g}-{\rm r} \gtsm 1.6$).
Figure 7 shows the color--magnitude 
diagrams for this cluster. The morphological 
separation between stars and galaxies in this case is more difficult than in 
Cl~1613+3104 because the seeing and the linear sampling are poorer (Paper I). 
For this reason, we do not introduce a discrimination {\sl a priori} 
between stars, galaxies and unidentified objects, neither in the cluster 
field nor in the comparison field.
The E/S0 color--magnitude sequence is barely visible in this cluster.

\begin{figure}
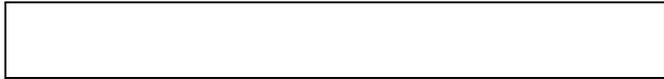
 
%\picplace{8cm}
\picplace{1cm}
%\centerline{\psfig{figure=fig7.ps,height=11.6cm}}
\caption [] {Color--r magnitude diagrams in the field of Cl~1600 +4109 :
{\bf a} $({\rm B}-{\rm r})$, {\bf b} (${\rm g}-{\rm r}$).
The (B $-$ r)(r) locus of B$_{\rm com}$ and the 
(g $-$ r)(r) locus of g$_{\rm com}$
are given by solid lines, whereas r$_{\rm com}$ is shown by a dashed line.
Galaxies are plotted as black dots, whereas stars and
unidentified objects appear as stars and plus signs, respectively
}
\end{figure}

Table 2 shows for different samples the excess 
of objects in the cluster field with respect to an empty field. 
Figure 8 shows the color--color diagrams of the 
cluster and comparison fields per bin in r magnitude. 
The importance of the blue population of galaxies in the color-color 
diagrams increases with magnitude (as for Cl~1613+3104). Note that in this case
the same warning comment as in Cl~1613+3104 is applicable concerning 
the absolute distribution in morphological types. 

\begin{table}
\caption[]{Distribution of the excess of galaxies in the field of
Cl~1600+4109 with respect to the comparison field.}
\begin{flushleft}
\begin{tabular}{lrrrrrl}
\hline\noalign{\smallskip}
%\halign{#\hfil&&\quad#\hfil\cr
%\noalign{\hrule\medskip}
 & E/S0 & S/Im & EB \\
\noalign{\smallskip}
\hline\noalign{\smallskip}
%\noalign{\medskip\hrule\medskip}
(B $-$ r) &$ [2.2, 3.3] $&$ [0.6, 2.2] $&$ < 0.6$\\
(g $-$ r) &$ [1.1, 2.3] $&$ [0.4, 1.6] $&$ < 1.0$\\
\noalign{\smallskip}
\hline\noalign{\smallskip}
%\noalign{\medskip\hrule\medskip}
All & 25\% & 75\% & 0\% \\
$22\leq{\rm r}\leq24$ & 13\% & 87\% & 0\% \\
${\rm r}\geq24$ & 0\% & 17\% & 83\% \\
\noalign{\smallskip}
\hline
%\noalign{\medskip\hrule\medskip}}
\end{tabular}
\end{flushleft}
\end{table}

   According to the definition of blue population by BO, the lack
of E/S0 galaxies in this cluster gives a value of f$_{\rm B}$ close to
1. The number of galaxies with M$_{\rm V}<-20$ (equivalent to 
${\rm r}<22.7$ for elliptical 
galaxies at $z = 0.540$) is only 70 within the cutoff radius of the cluster
(Sect.\ 5.2), and 10 within R$_{30}$ (= 26$\arcsec$, see Sect.\ 5.2). 
The contamination level, as estimated from the comparison
field, is close to $10 \%$. Almost all these objects are at least 0.2
mag bluer than a typical E galaxy in B $-$ g and g $-$ r, so the fraction
of blue objects inferred is very high: f$_{\rm B} = 0.8\pm0.5$. The result is
the same when the sample is limited to galaxies within the completeness
magnitude in g (g $\leq 24.5$).
In this case there are 10 galaxies within
R$_{30}$, 9 of which are ``blue'', with a contamination level close to
$50\%$.

\begin{figure}
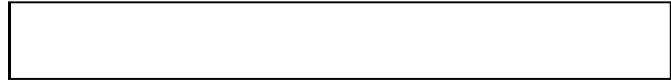

%\picplace{11.5cm}
\picplace{1cm}
%\centerline{\psfig{figure=fig8.ps,height=11.6cm}}
\caption [] {Color--color diagrams in the field of Cl~1600+4109 and
in the comparison field. Three magnitude 
intervals are considered: {\bf a and b} ${\rm r} \leq 22$, {\bf c and d} 
$22 <{\rm  r} < 24$, {\bf e and f} ${\rm r} \geq 24$. 
Objects are identified as in Fig.\ 7.
The colors predicted for
the different types of galaxies at $z=0.540$, from E to Im, 
and the typical error boxes in each magnitude bin are also
shown
} 
\end{figure}

\subsection{ The 2D distribution of galaxies }

The shape of Cl~1600+4109 is obtained 
from the projected number density of galaxies.
Figure 9 shows the isocontour map. The ellipticity of the cluster is
$\epsilon\simeq0.75$ and the position angle $\theta\simeq45\degr$. 
The cluster is dominated by 
a small group of galaxies in its central part (see Fig.\ 2 in Paper I). There
is a bright galaxy of magnitude r = 19.58 
in the center of the group, with colors
fully compatible with those expected for an elliptical galaxy at $z=0.540$. 

   The cluster profile and the cutoff radius have been calculated 
as in Cl~1613+3104, using the projected density of the comparison field.
The cutoff radius is $78\arcsec$ (635~h$^{-1}$~kpc); the total 
number of objects within it is 137, and there are 100 objects in the 
comparison field, under the same conditions. 
most objects in excess with respect to the empty field are found beyond the 
completeness limit in magnitude in the three filters. 
We have obtained an estimate of the concentration parameters without 
introducing any selection in magnitude. The concentration parameter is 
${\rm  C}=0.40$, with ${\rm R}_{30}=26\arcsec$ (212 h$^{-1}$ kpc) and
N$_{30}=10$. The central galaxy density, ${\rm N}_{0.5}$, is strongly dependent
on the filter used. The highest value is attained in the B filter, with
${\rm N}_{0.5} = 20$, as expected for a so blue population of galaxies.
The corresponding velocity dispersion 
(Bahcall 1981) is about 800--1000~km~s$^{-1}$. 
These values of ${\rm N}_{0.5}$
and C should only be taken as a rough estimate, because the sample has not been
limited in magnitude.
The projected galaxy density profile, 
according to Eq.\ (2) with $\alpha\equiv1$, is:
$\sigma_0=0.008\pm0.002~{\rm gal~arcsec}^{-2}$
and R$_{\rm c}=12\arcsec\pm8\arcsec\,
(100\pm70~{\rm h}^{-1}~{\rm kpc})$.

\begin{figure}
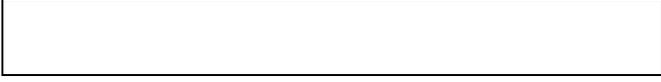
 
%\picplace{14.5cm}
\picplace{1cm}
%\centerline{\psfig{figure=fig9.ps,height=11.6cm}}
\caption [] {Isocontour map of the projected number density of galaxies in the
field of Cl~1600+4109. The mean projected density over the cluster field is 
$5.0\times10^{-3} {\rm objects~arcsec^{-2}}$, and the difference between 
two successive contour levels is 0.5 times the mean density, starting at 1.
X and Y coordinates are in arcsec, with the center of the cluster at
($80\arcsec$, $130\arcsec$) }
\end{figure}

The $\lambda_{\rm n}$-method does not detect any
obvious segregation either in magnitude or in color.
There is a slight tendency among red objects to be more concentrated
than blue objects, but this trend is of the same order as the errors.

\section{ Discussion and conclusions}
Cl~1613+3104 is a rich, concentrated cluster, dominated 
by a compact group of luminous 
galaxies within a radius of 121 h$^{-1}$ kpc around the center. The cluster 
shows strong evidence for segregation in luminosity and color, and both
effects seem to be independent.
%% as we find a segregation in color when the sample is 
%% limited within a magnitude bin, and a segregation in magnitude 
%% when the sample is selected within a narrow color interval. 
Red and bright objects tend to concentrate around the center of the cluster. 
The distribution of galaxies in colors shows two
different populations of objects. The red population
is compatible with the colors expected for E/S0 
type galaxies at the cluster redshift. These objects dominate the population of
cluster galaxies at low magnitudes and they are located preferentially in the 
central part of the cluster. The blue population is explained in part by the 
presence of normal star-forming S/Im galaxies, but a population of galaxies 
bluer than normal Im galaxies at the cluster redshift is also detected.
Moreover, the 2D distribution for the red population
is clearly more concentrated around the center of the cluster
than the blue one.

The detection level in Cl~1600+4109 is lower than in Cl~1613+3104,
and most cluster galaxies are found beyond the completeness limits in magnitude.
Nevertheless, its concentration and richness parameters indicate that
we are probably dealing with a rich, concentrated cluster. It
is dominated by a bright galaxy in its 
central region, with colors fully compatible
with those expected for E galaxies at this redshift. The fraction of E/S0
galaxies is low in this cluster (less than $15 \%$ within the completeness
limit in r), and the population of galaxies is dominated by S/Im
galaxies. There is no evidence of segregation in magnitude or in color.

   In both clusters, the luminosity function in B is compatible with the
standard values derived by Colless (1989) for a sample of rich low-redshift
clusters. The absolute magnitudes derived for their brightest galaxies,
after applying the standard k+e correction, are similar. The main
galaxy in Cl~1600+4109 is slightly brighter than that of Cl~1613+3104
(${\rm M}_{\rm r} = -23.40$ compared to ${\rm M}_{\rm r} = -23.01$,
within the isophote corresponding to $1\sigma$ of the background sky noise).
Comparison of the two histograms in M$_{\rm B}$ for Cl 1600+4109 and 
Cl~1613+3104, after
correction through the comparison field (Figs.\ 1 and 6), shows that
the completeness magnitude in Cl~1613+3104 
is 1.2 mag fainter than in Cl~1600+4109
when a k correction for E/S0 galaxies is used, and this value reduces to
0.9 mag when a standard k+e correction is applied. Nevertheless, most galaxies
with magnitudes around B$_{\rm com}$ in both clusters are S/Im galaxies. When 
the k correction used is that of an Sb galaxy, the completeness magnitude in
Cl~1613+3104 is 0.6 mag fainter than in Cl~1600+4109, and the difference
reduces to 0.3 mag after a k+e correction. So the absolute detection level 
in Cl~1600+4109 is poorer than in Cl~1613+3104. The main reason for this is
probably the difference in the seeing conditions between both clusters. 
Cl~1613+3104 was observed with an averaged seeing of $\sim 1\farcs3$,
whereas the seeing was $\sim 2\farcs1$ for Cl~1600+4109, but the
isophotes used to integrate fluxes were 0.5 mag fainter in the later
(see Paper I). A straightforward simulation of photometry has been done, 
taking the best r image of Cl~1613+3104 (seeing $= 1\farcs1$) and 
debasing it to an equivalent seeing of $\sim 2\farcs5$. As expected, a
general tendency appears in the isophotal magnitudes: objects in the 
bad-seeing image tend to be measured as fainter than in the 
good-seeing one, and this effect increases with the magnitude. The brightest 
objects (${\rm r} \ltsm 20$), such as the main cluster-galaxies, are almost
unaffected, the difference being lower than 0.1 mag. This difference 
is about 0.2 mag for objects with ${\rm r} \simeq 23$, and it increases
to 0.4 mag for the faintest objects in the sample, up to ${\rm r} = 25$.
Using a fainter isophote to integrate magnitudes, as it was done 
for Cl~1600+4109, strongly reduces the effect. When the isophotal magnitude 
is the same in both the good and the bad seeing images, the
difference in magnitude for the faintest objects can be as large as 0.7 to
1 mag. 

\begin{table}
\caption[]{Expected colors for a single burst of star 
formation as a function of its age, as seen at the redshifts of both clusters}
\begin{flushleft}
\begin{tabular}{lrrrl}
\hline\noalign{\smallskip}
%\halign{#\hfil&&\quad#\hfil\cr
%\noalign{\hrule\medskip}
Age       &{\rm B} $-$ {\rm r} & {\rm g} $-$ {\rm r}&  {\rm B} $-$ {\rm r} & 
{\rm g} $-$ {\rm r} \\
$\left[ {\rm log(yr)} \right]$     & \multispan2 [{\it z}~=~0.415] & 
\multispan2 [{\it z}~=~0.540]  \\
\noalign{\smallskip}
\hline\noalign{\smallskip}
%\noalign{\medskip\hrule\medskip}
     5   &         $-$0.10     &       0.04   &           $-$0.06    & 0.08 \\
     6   &         $-$0.13     &       0.02   &           $-$0.09    & 0.06 \\
     7   &          0.32       &       0.40   &            0.27      & 0.36 \\
     8   &          0.85       &       0.83   &            0.87      & 0.86 \\ 
     9   &          1.44       &       1.30   &            1.51      & 1.36 \\
\noalign{\smallskip}
\hline
\end{tabular}
\end{flushleft}
\end{table}
%\noalign{\medskip\hrule\medskip}}

In Cl~1613+3104, the population of galaxies is dominated by the E/S0 type,
at least up to the completeness limit in r. 
%% The population of normal S/Im 
%% galaxies represents about $33\%$ of the sample, 
%% and the remaining $23\%$ of galaxies
%% are extremely blue objects, with colors 
%% bluer than those expected for Im galaxies.
In the case of Cl~1600+4109, there is a lack of
E/S0 galaxies (they are less than $15 \%$ of the sample) and the population
is mainly composed of S/Im galaxies up the completeness limit in r. 
The extremely blue population dominates the sample at faint magnitudes 
(${\rm r} > 24$) in both clusters.
The general trend is that the fraction of blue galaxies in
excess with respect to an empty field
increases with magnitude. Nevertheless, getting the absolute distribution in
morphological types is only possible when the sample of galaxies is 
limited within the magnitude of completeness in {\sl all} the filters
involved. Otherwise, a bias is introduced, as can be seen 
in Figs.\ 2 and 7. The result is that the fraction of blue objects detected 
in both clusters should be considered as an upper bound because the detection of
red objects is not complete. In Cl~1613+3104, the color-color diagrams are 
only complete for objects bluer than B $-$ r = 0.25 
and g $-$ r = 0.5 in the interval 
$22 \leq {\rm r} \leq 24$ and B $-$ r = $-$0.5 and g $-$ r = $-$0.25 when
${\rm r} \geq 24$. In Cl~1600+4109, this condition 
is fulfilled for objects bluer
than B $-$ r = 0.75 and g $-$ r = 0.75 if $22 \leq {\rm r} \leq 24$, and 
B $-$ r = $-$0.25 and g $-$ r = $-$0.25 when ${\rm r} \geq 24$. 
However, even if the absolute ratio of blue to red objects is biased, 
the both cluster fields show clearly 
an excess of extremely blue objects in comparison with 
an empty field, and the number of such objects increases with the magnitude.
Table 3 gives the location in the color-color plane of a single burst of
star formation, as seen at the redshifts of both clusters, as a function 
of its age. It is assumed that the light coming from the stellar mass 
created by the burst is responsible for the bulk of the energy
in the spectra. When we compare these values with the 
color-color distributions shown in Figs.\ 3 and 8, 
it appears that most extremely blue objects in both clusters 
are compatible with the colors expected for such burst 
models younger than a few $10^{7}$~yr. High quality images of these clusters 
could confirm the hypothesis that
these galaxies are going through mergers or tidal interactions. These
observations would reinforce the idea that triggers of starburst processes
were more frequent in the past than in the present epoch.

\acknowledgements 
We wish to  thank the technical staff  of the CAHA for  his
help  during  the  photometric  runs  at  the  3.5~m  telescope.  We are
grateful to J.-F.  Le Borgne for his help during the data reduction.   We
wish to thank  G. Bru\-zual  for allowing the use of his code  for the
spectral evolution  of galaxies.   This  work has  been supported by the
Spanish DGICYT program PB90-0448 of the {\it Mi\-nis\-te\-rio de 
Edu\-ca\-ci\'on y
Cien\-cia} and, partially, by the Human Capital and Mobility Programme (EU), 
under contract CHRX-CT92-0044.

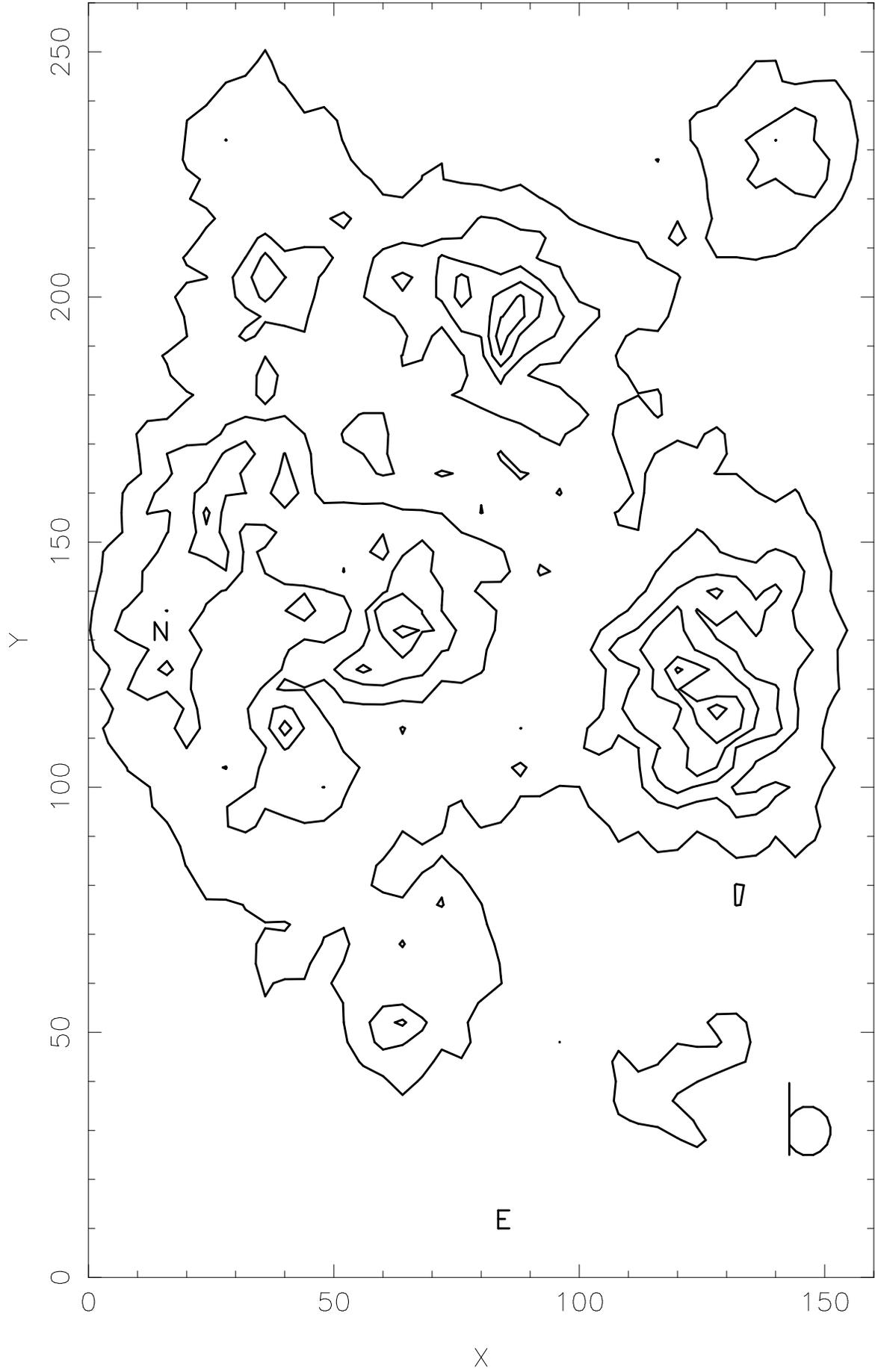
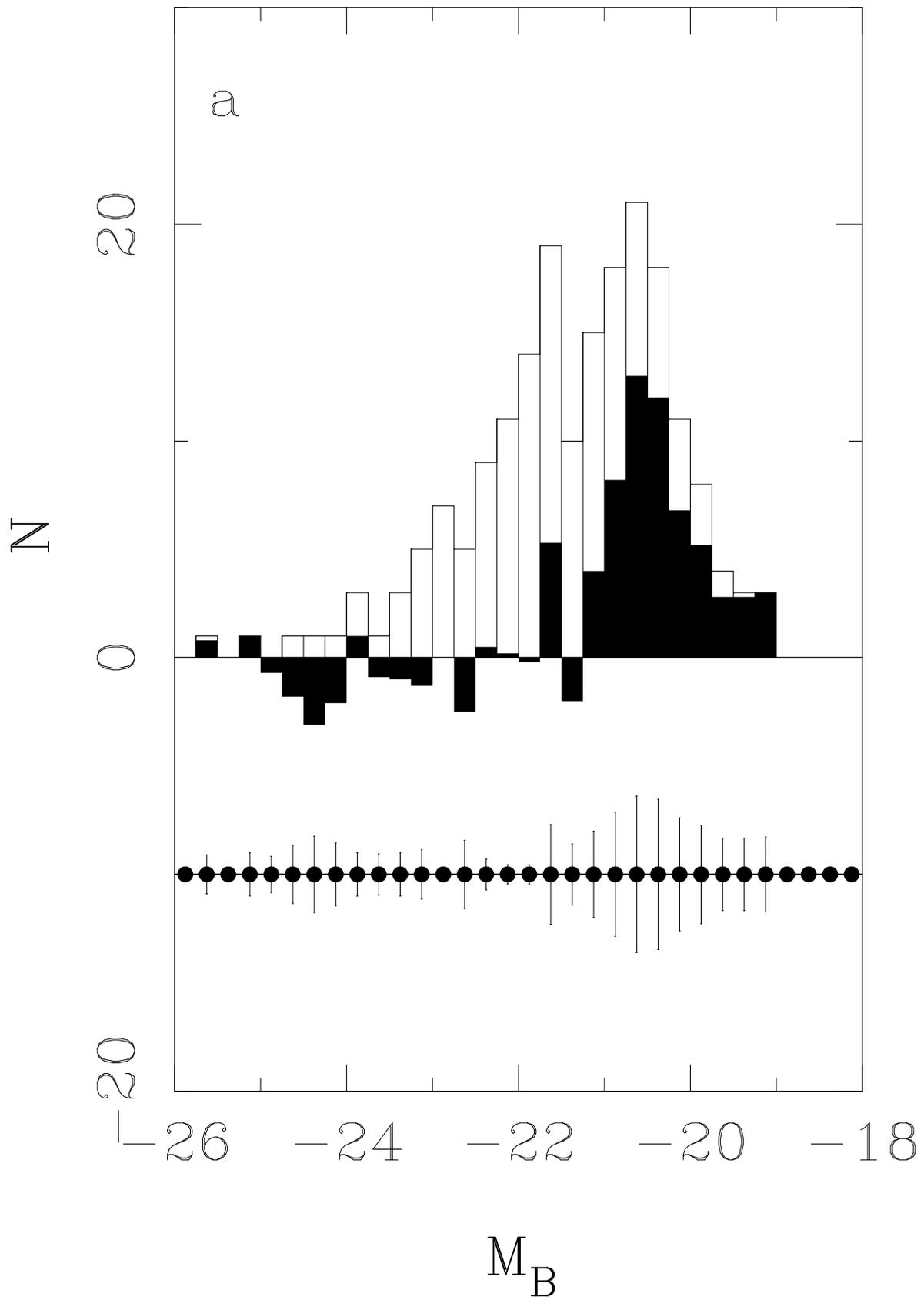
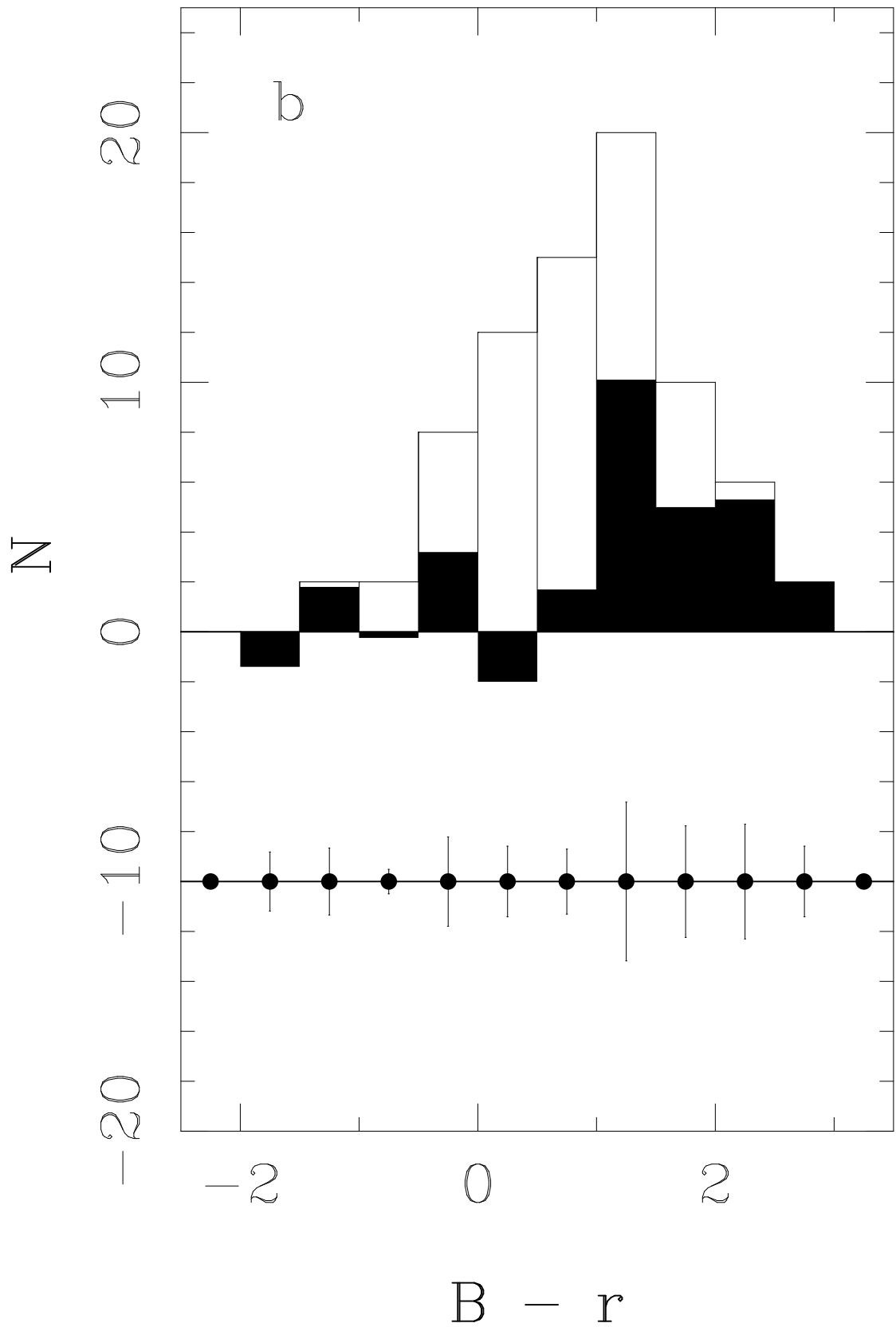
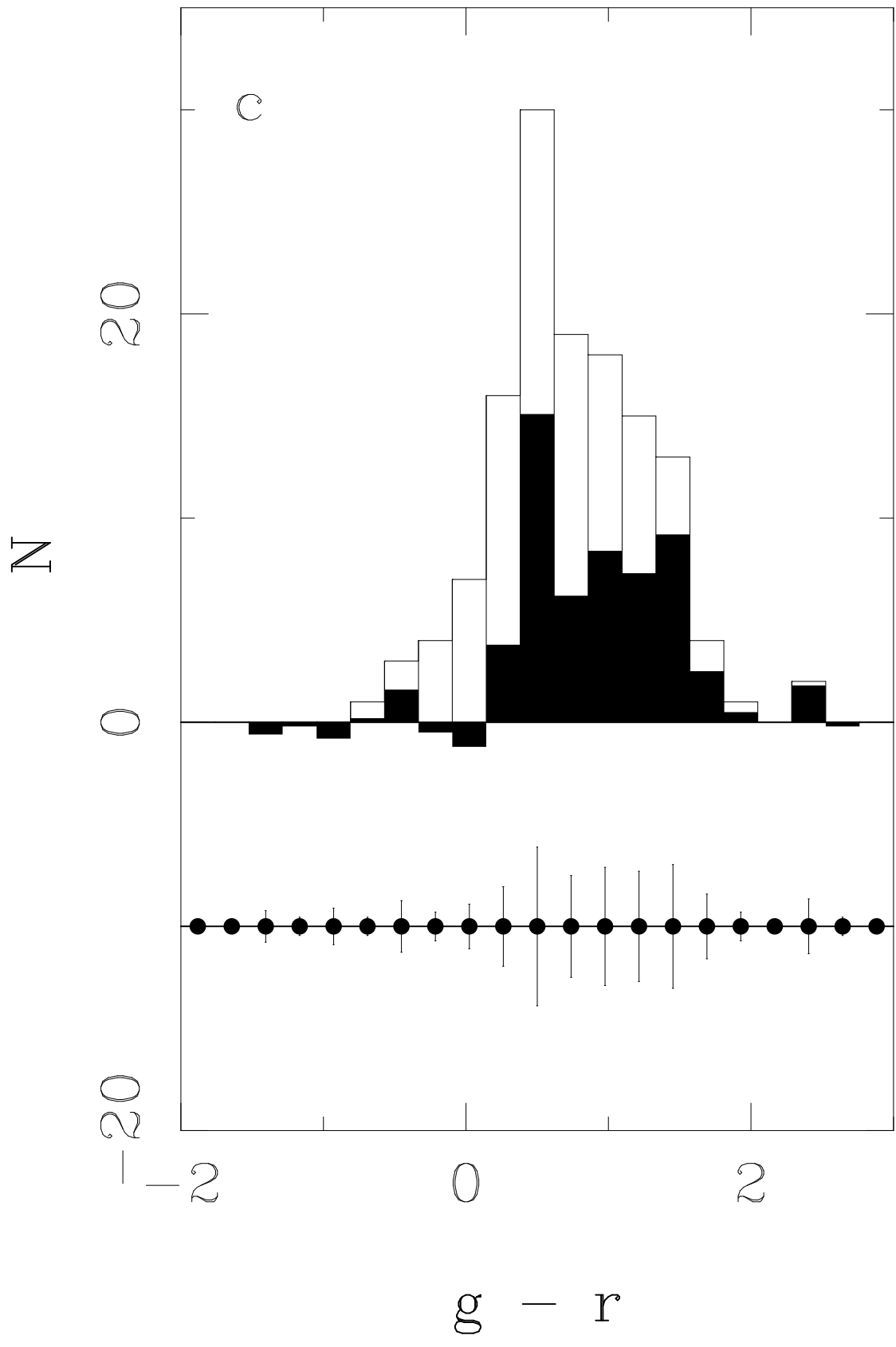
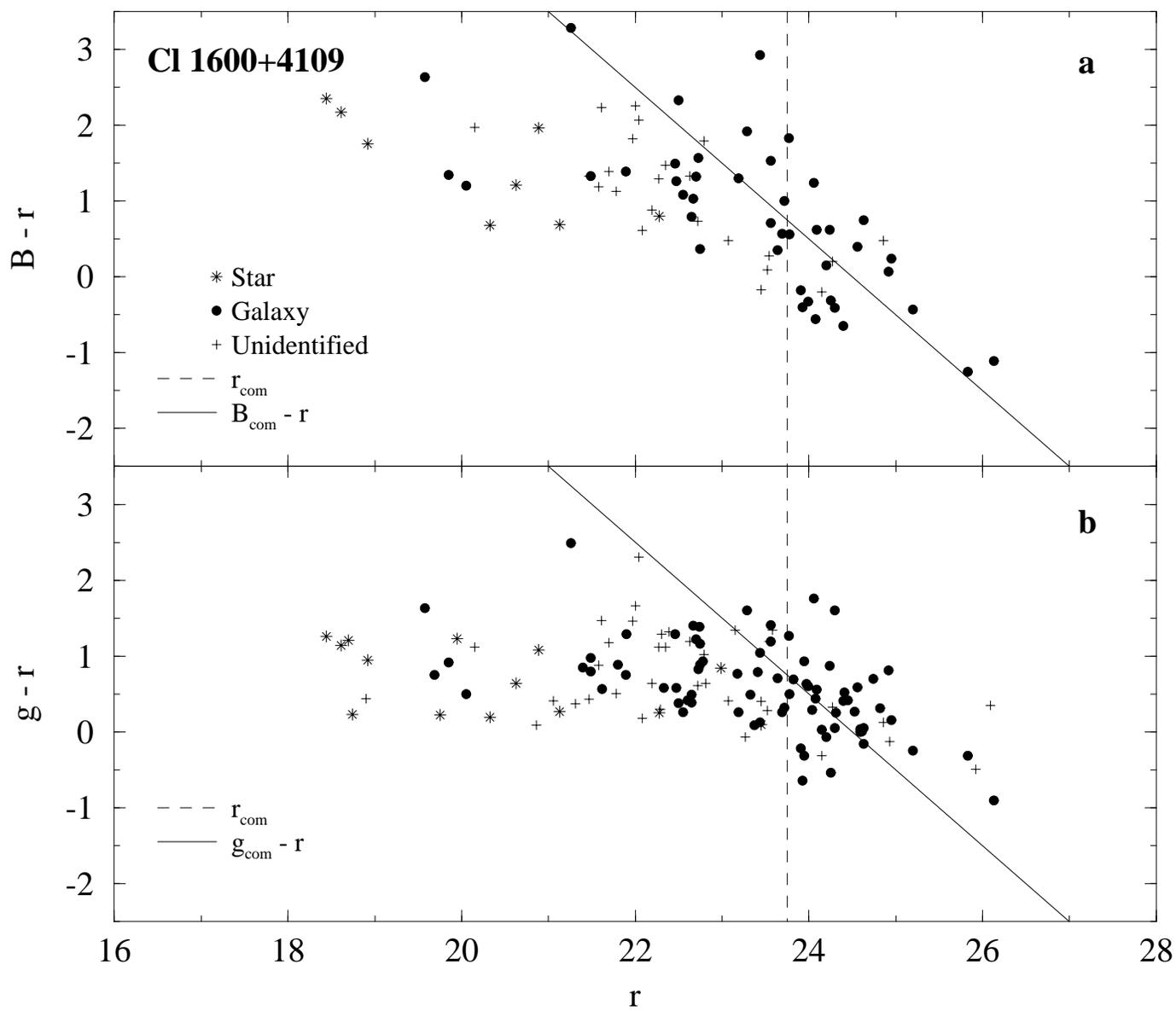
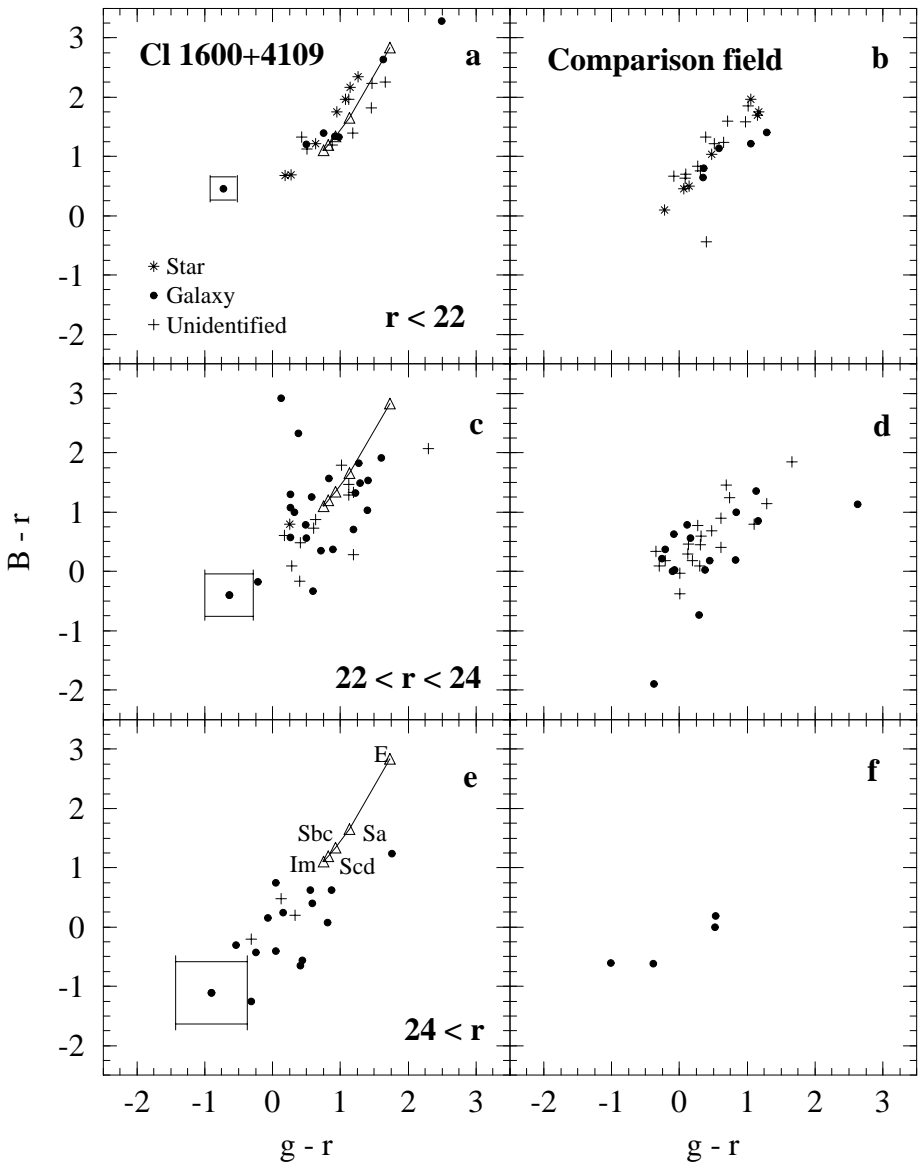
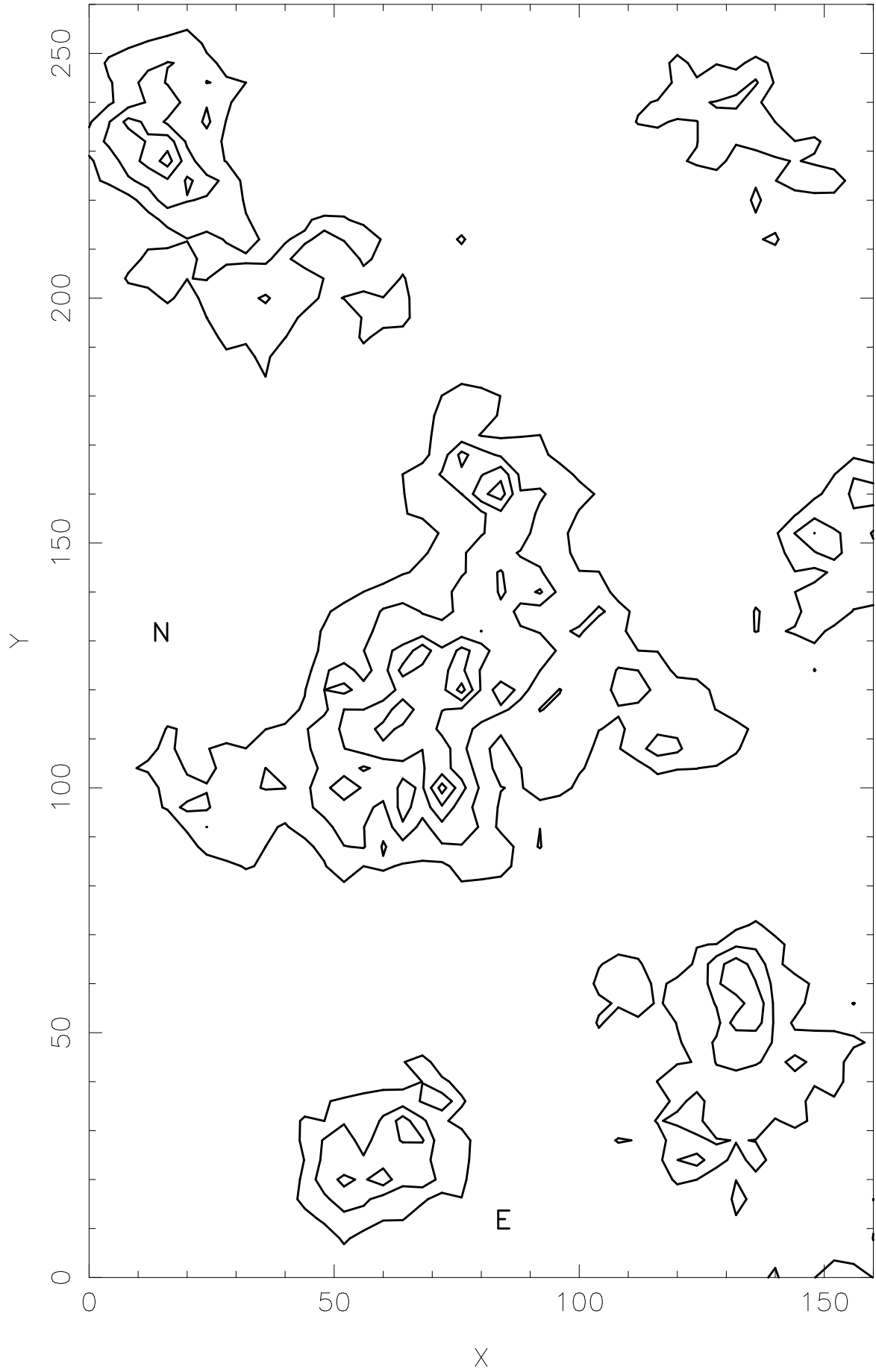
\end{document}